\documentstyle[12pt]{article}

\newcommand{\wh}{\widehat}
\newcommand{\wt}{\widetilde}
\newcommand{\U}{{\cal U}}
\newcommand{\T}{{\cal T}}
\newcommand{\SS}{{\cal S}}
\newcommand{\V}{{\cal V}}
\newcommand{\OO}{{\cal O}}
\newcommand{\bm}{{\bar m}}
\newcommand{\bn}{{\bar n}}
\newcommand{\bG}{{\bar G}}
\newcommand{\bB}{{\bar B}}
\newcommand{\bH}{{\bar H}}
\newcommand{\bQ}{{\bar Q}}
\newcommand{\bp}{{\bar p}}
\newcommand{\bC}{{\bar C}}
\newcommand{\bK}{{\bar K}}
\newcommand{\bM}{{\bar M}}
\newcommand{\bF}{{\bar F}}
\newcommand{\ba}{{\bar a}}
\newcommand{\bb}{{\bar b}}
\newcommand{\bpsi}{{\bar \psi}}
\newcommand{\bchi}{{\bar \chi}}
\newcommand{\brho}{{\bar \rho}}
\newcommand{\bL}{{\bar L}}
\newcommand{\bA}{{\bar A}}
\newcommand{\bq}{{\bar q}}
\newcommand{\bg}{{\bar g}}
\newcommand{\bPhi}{{\bar\Phi}}
\newcommand{\bmu}{{\bar\mu}}
\newcommand{\bnu}{{\bar\nu}}
\newcommand{\be}{\begin{equation}}
\newcommand{\ee}{\end{equation}}
\newcommand{\ben}{\begin{eqnarray}\displaystyle}
\newcommand{\een}{\end{eqnarray}}

\newcommand{\refb}[1]{(\ref{#1})}
\newcommand{\p}{\partial}
\newcommand{\ten}{{(10)}}
\newcommand{\two}{{(2)}}

\newcommand{\sectiono}[1]{\section{#1}\setcounter{equation}{0}}

\title{DUALITY SYMMETRY GROUP OF TWO DIMENSIONAL HETEROTIC
STRING THEORY}

\author{Ashoke Sen \\
Tata Institute of Fundamental Research \\
Homi Bhabha Road, Bombay 400005, INDIA \\
sen@theory.tifr.res.in, sen@tifrvax.bitnet
}

\begin{document}

\maketitle

\begin{abstract}

The equations of motion of the massless sector of the two dimensional
string theory, obtained by compactifying the heterotic string theory
on an eight dimensional torus, is known to have an affine $o(8,24)$
symmetry algebra generating an $O(8,24)$ loop group.
In this paper we study how various known discrete $S$- and
$T$- duality symmetries of the theory are embedded in this
loop group. This allows us to identify the generators of the discrete
duality symmetry group of the two dimensional string theory.

\end{abstract}

\vfill

\vbox{\hbox{TIFR-TH-95-10}\hbox{hep-th/9503057}\hbox{March, 1995}}
\hfill ~

\sectiono{Introduction and Summary}

There is mounting evidence that heterotic string theory compactified
on a six dimensional torus is invariant under an $SL(2,Z)$ group of
duality transformation that interchanges the strong and weak coupling
limits of the
theory\cite{MONTOLIVE}-\cite{BERVAF}.
This duality symmetry is known as $S$-duality,
and is distinct from the target space duality or $T$-duality
symmetry\cite{ROCVER,TREV}
of the theory which generates the group $O(6,22;Z)$.
$T$-duality symmetry holds order by order in string perturbation theory,
and hence is easy to test, whereas $S$-duality transformation mixes
different orders in string perturbation theory and hence is not testable
in perturbation theory. Nevertheless, non-trivial tests of
$S$-duality symmetry have been designed, and so far the results of all
such tests support the hypothesis that $S$-duality is a genuine symmetry
of the full string theory. If this hypothesis turns out to be correct,
then we can use this symmetry to unravel some of the non-perturbative
features of string theory, since this symmetry acts non-trivially on
the string coupling constant. In particular, we can calculate scattering
amplitude of elementary string states at strong coupling by
calculating the scattering amplitude of solitons at weak coupling.

Although string theory in four dimensions is the theory that is of most
interest to us, the particular string theory in which $S$-duality is
most manifest, namely, ten dimensional heterotic string theory compactified
on a torus, does not give a phenomenologically acceptable four dimensional
theory. There are of course four dimensional theories which are much
better for low energy phenomenology (including models which have $SU(3)
\times SU(2) \times U(1)$ as low energy gauge group and three
generations of quarks and leptons\cite{KACHRU}), but at present
it seems to be difficult to understand the role of $S$-duality in such
theories. (Some progress in this direction has been made in
refs.\cite{SEIWIT}-\cite{SEIBERG}.) A
more fruitful approach seems to be not to restrict ourselves only to
physically motivated models, but study $S$-duality in a more general
context. One might hope that understanding $S$-duality in a widely
different class of models may finally give us a clue as to how it
manifests itself in a physically motivated string theory.

Guided by this principle, in a previous paper we had studied the
manifestation of $S$-duality in a three dimensional string theory, obtained
by compactifying ten dimensional heterotic string theory on a seven
dimensional torus\cite{THREED}.
It was found that in three dimensions the $T$-duality
group $O(7,23;Z)$ and the $S$-duality group combine into a much bigger
group\break
\noindent $O(8,24;Z)$. This analysis was facilitated by an earlier work of
Marcus and Schwarz\cite{MARSCH},
where they had studied the symmetries of the
three dimensional supergravity theory, which forms the massless sector
of the three dimensional string theory, and had found it to be
$O(8,24)$. When we take into account the massive string states, only
a discrete subgroup $O(8,24;Z)$ of $O(8,24)$ has a chance of being an
exact symmetry of the spectrum, and this was conjectured to be the
symmetry of the three dimensional string theory.
One support to this conjecture comes from the fact
that this discrete $O(8,24;Z)$ subgroup is generated by the $T$-duality
group $O(7,23;Z)$ of the three dimensional string theory, and the
$S$-duality group $SL(2,Z)$ of the four
dimensional theory before compactification from four to three dimensions.
Thus if $S$-duality is a symmetry of the four dimensional theory, and if
it survives compactification of one of the dimensions, then $O(8,24;Z)$
must be the symmetry group of the three dimensional theory. (Similar
arguments have been used in ref.\cite{ASPIN} to construct the full
$T$-duality group of string theories compactied on $K3$.)

In this paper, we shall carry out a similar analysis for the two
dimensional string theory, obtained by compactifying heterotic string
theory on an eight dimensional torus. Again our analysis is facilitated
by known results about two dimensional supergravity theory, whose
classical equations of motion are known to be invariant under an
$\wh {o(8,24)}$ current algebra
symmetry\cite{GEROCH}-\cite{SCHTWOD}. (For a history of this subject,
see refs.\cite{NICOLAI,SCHTWOD}.)
Thus one would expect the
duality symmetry group of the two dimensional string theory to be a
discrete subgroup of the loop group $\wh{O(8,24)}$. (This was already
conjectured by Hull and Townsend\cite{HULL}).
If we assume that the $S$-duality is a symmetry of the theory even after
two of the four dimensions are compactified, then the minimal
duality symmetry group of the two dimensional theory
is generated by the usual $T$-duality
transformations $O(8,24;Z)$
of the two dimensional string theory, and
the $S$-duality transformations of the four dimensional string theory.
As in the case of three dimensional string theory, the $S$- and $T$-
duality transformations do not commute, and generate a much bigger
group. Following the convention of Hull and Townsend\cite{HULL},
we shall call
this the $U$-duality group, and denote this by $G$.

The main aim of this paper will be to find explicit
representation of the generators of the group $G$. As we shall see,
these will be represented by O(8,24) valued functions (possibly
containing poles at the origin) of a real variable $v$.
There are two
main steps involved in this project. In section 2 we study how the
various known symmetry groups of two and higher dimensional supergravity
theories are embedded in the loop group $\wh{O(8,24)}$. This includes
the manifest $O(8,24)$ symmetry group of the two dimensional supergravity
theory that contains the $T$-duality group $O(8,24;Z)$ as its natural
subgroup, as well as the $O(8,24)$ symmetry of the three dimensional
supergravity theory that contains the $S$-duality group $SL(2,Z)$ as
its subgroup. Thus this analysis provides an explicit embedding
of the various known duality transformations in the loop group
$\wh{O(8,24)}$.  In section 3 we
use this knowledge to construct explicitly the elements of the loop
group $\wh{O(8,24)}$ (represented by $O(8,24)$ matrix valued functions
of a real variable) corresponding to the generators of the duality
transformation.
The appendix contains some of the technical results that
are needed for the analysis of section 2.

We conclude this section by stating some of the conventions that
we shall be using in this paper. We use the normalization
$\alpha'=16$. We shall denote algebras by lower case letters
{\it e.g.} $o(8,24)$, and groups by upper case
letters {\it e.g.} $O(8,24)$. We shall use
a hat to denote the affine algebra or loop group, {\it e.g.}
$\wh{o(8,24)}$
or $\wh{O(8,24)}$. Algebras or groups without hat will denote the
ordinary Lie algebras or Lie groups generated by the zero
mode subalgebra of
the corresponding affine algebra. Finally, in our convention, the group
element $g$ and the generators $T^\alpha$ are related by
$g=\exp(\theta_\alpha T^\alpha)$; note the absence of $i$ in the exponent.

\sectiono{Embedding of Various $o(8,24)$ Subalgebras in the Affine Algebra
$\wh{o(8,24)}$}

We start with the massless sector of the heterotic string
theory in ten dimensions. This theory is described by the action
\ben \label{e1}
S &=& {2\pi \over (8\pi)^8}
\int d^{10}z \sqrt{-G^\ten} e^{-\Phi^\ten} \Big[R^\ten +
G^{\ten MN} \p_M \Phi^\ten
\p_N\Phi^\ten \nonumber \\
&& -{1\over 12} H^\ten_{MNP} H^{\ten MNP}
-{1\over 4} F^{\ten I}_{MN} F^{\ten IMN}\Big]\, ,
\een
where $R^\ten$ denotes the scalar curvature,
\ben \label{e1a}
F^{\ten I}_{MN} &=& \p_M A^{\ten I}_N - \p_N A^{\ten I}_M, \nonumber \\
H^\ten_{MNP} &=& (\p_M B^\ten_{NP} -{1\over 2}
A^{\ten I}_M F^{\ten I}_{NP})
+\hbox{cyclic permutations of $M$, $N$, $P$}\, , \nonumber \\
\een
and $G^\ten_{MN}$, $B^\ten_{MN}$, $A^{\ten I}_M$ and $\Phi^\ten$
($0\le M\le 9$) denote the metric, anti-symmetric tensor field,
$U(1)^{16}$ gauge fields, and the dilaton field in ten dimensions.
$z^M$ are the coordinates of the ten dimensional space.
The overall normalization factor multiplying the action has been chosen
appropriately for later convenience.
We now dimensionally reduce the theory to two dimensions by
defining\cite{FERRARA,DIMRED},
\ben \label{e2}
z^\mu & = & x^\mu, \quad 0\le \mu \le 1 \nonumber \\
z^{m+1} & = & y^m, \quad 1\le m \le 8\, .
\een
We shall take each of the coordinates $y^m$ to represent
a compact direction with length $8\pi$ : $y^m\equiv y^m+8\pi$.
We take various fields to be independent of the coordinates $y^m$ and
define,\footnote{The last term in the expression for $B_{\mu\nu}$
was inadvertently left out in refs.\cite{SREV,THREED}.}
\ben \label{e3}
&& \wh G_{mn}  = G^\ten_{m+1,n+1}, \quad  \wh B_{mn}  =
B^\ten_{m+1, n+1}, \quad  \wh A^I_m  = A^{\ten I}_{m+1},
\nonumber \\
&& A^{(2m-1)}_\mu  = {1\over 2}\wh G^{mn} G^\ten_{n+1,\mu}, \quad
A^{(I+16)}_\mu = -({1\over 2} A^{\ten I}_\mu - \wh A^I_n
A^{(2n-1)}_\mu), \nonumber \\
&&  A^{(2m)}_\mu = {1\over 2}
B^\ten_{(m+1)\mu} - \wh B_{mn} A^{(2n-1)}_\mu + {1\over 2}\wh A^I_m
A^{(I+16)}_\mu, \nonumber \\
&& G_{\mu\nu} = G^\ten_{\mu\nu} - G^\ten_{(m+1)\mu} G^\ten_{(n+1)\nu} \wh
G^{mn}, \nonumber \\
&& B_{\mu\nu} = B^\ten_{\mu\nu} - 4\wh B_{mn} A^{(2m-1)}_\mu
A^{(2n-1)}_\nu - 2 (A^{(2m-1)}_\mu A^{(2m)}_\nu - A^{(2m-1)}_\nu
A^{(2m)}_\mu) \nonumber \\
&& \qquad \qquad  +2 \wh A^I_n (A^{(2n-1)}_\mu A^{(I+16)}_\nu
 - A^{(2n-1)}_\nu A^{(I+16)}_\mu)\, ,
\nonumber \\
&& \Phi = \Phi^\ten - {1\over 2} \ln\det \wh G, \quad \quad
\quad 1\le m, n \le 8, \quad
0\le \mu, \nu \le 1, \quad 1\le I \le 16. \nonumber \\
\een
Here $\wh G^{mn}$ denotes the inverse of the matrix $\wh G_{mn}$.
We also define,
\be \label{e4}
\wh C_{mn} = {1\over 2} \wh A^I_m \wh A^I_n\, .
\ee
For every $m,n$ ($1\le m,n \le 8$) we define $H_{mn}$ to be the
$2\times 2$ matrix
\be \label{e5}
H_{mn} = \pmatrix{\wh G^{mn} & \wh G^{mp}(\wh B_{pn} + \wh C_{pn})\cr\cr
(-\wh B_{mp} + \wh C_{mp}) \wh G^{pn} & (\wh G - \wh B + \wh C)_{mp}
\wh G^{pq} (\wh G + \wh B + \wh C)_{qn} \cr}\, ,
\ee
and for every $m$, $I$ ($1\le m\le 8$, $1\le I\le 16$) we define
$Q^{(I)}_m$ to be the two dimensional column vector
\be \label{e6}
Q^{(I)}_m = \pmatrix{\wh G^{mn} \wh A^I_n \cr\cr
(\wh G - \wh B + \wh C)_{mn} \wh G^{np} \wh A^I_p \cr}\, .
\ee
We also define $K$ to be a $16\times 16$ matrix whose components are,
\be \label{e7}
K_{IJ} = \delta_{IJ} + \wh A^I_m \wh G^{mn} \wh A^J_n \, .
\ee
In terms of $H$, $Q$ and $K$, we now define a $32\times 32$ matrix $M$ as
follows:
\be \label{e8}
M = \pmatrix{ H_{11} & \cdot & H_{18} & Q^{(1)}_1 & \cdot & Q^{(16)}_1 \cr
              \cdot  & \cdot & \cdot  & \cdot     & \cdot & \cdot      \cr
           H_{81} & \cdot & H_{88} & Q^{(1)}_8 & \cdot & Q^{(16)}_8 \cr\cr
        Q^{(1)  T}_1 & \cdot & Q^{(1) T}_8 &      &       &            \cr
             \cdot   & \cdot & \cdot  &           &   K   &            \cr
        Q^{(16) T}_1 & \cdot & Q^{(16) T}_8 &     &       &            \cr
}\, .
\ee
$M$ satisfies
\be \label{e9}
M^T = M, \qquad M L M^T = L\, ,
\ee
where $L$ is a $32\times 32$ matrix, defined as,
\be \label{e10}
L = \pmatrix{\sigma_1 &&&& \cr &\cdot &&& \cr && \cdot && \cr
&&& \sigma_1 & \cr
&&&& -I_{16}\cr}\, , \qquad
\sigma_1 = \pmatrix{& 1 \cr 1 &\cr}\, .
\ee
$I_n$ denotes the $n\times n$ identity matrix.
It is well known that the matrix $M$ contains exactly the same information
as the set of fields $\{\wh G_{mn}, \wh B_{mn}, \wh A^I_m\}$.

We now use the fact that the antisymmetric tensor field $B_{\mu\nu}$ and
the gauge fields $A_\mu^{(a)}$ ($1\le a \le 32$) have no dynamics in two
dimensions. Thus we can set them to zero. The remaining dynamical
variables in two dimensions are given by the fields $\Phi$, the matrix $M$,
and the two dimensional metric $G_{\mu\nu}$. In terms of these fields the
ten dimensional action \refb{e1} takes the form:
\be \label{e11}
S = (2\pi) \int d^2 x \sqrt{-G}\,
e^{-\Phi} \, [ R_G + G^{\mu\nu} \p_\mu\Phi
\p_\nu\Phi +{1\over 8} G^{\mu\nu} Tr(\p_\mu M L \p_\nu M L)]\, .
\ee
This action gives rise to the following set of independent
equations of motion in the conformal gauge $G_{\mu\nu}=\lambda e^{2\Phi}
\eta_{\mu\nu}$:
\be \label{e14}
\eta^{\mu\nu} \p_\mu \p_\nu (e^{-\Phi}) =0\, ,
\ee
\be \label{e15}
\p_\mu (e^{-\Phi} \eta^{\mu\nu} \p_\nu M M^{-1}) =0 \, ,
\ee
and,
\be \label{e16}
\p_\pm (\ln \lambda) \p_\pm (e^{-\Phi}) = \p_\pm^2(e^{-\Phi}) -{1\over 4}
e^{-\Phi} \,  Tr(\p_\pm M L \p_\pm M L) \, ,
\ee
where $x^\pm$ are the light cone variables,
\be \label{light}
x^\pm ={1\over \sqrt 2} (x^0 \pm x^1)\, .
\ee
This system of equations has an $o(8,24)$ current
algebra symmetry\cite{GEROCH}-\cite{SCHTWOD} which we shall denote by
$\wh{o(8,24)}$. It is generated by conserved charges $J^\alpha_n$
($-\infty <n <\infty$) and the central charge\cite{JULIA} $C$:
\be \label{e16a}
[J_m^\alpha, J_n^\beta]=f^{\alpha\beta\gamma}\eta_{\gamma\delta}
J^\delta_{m+n} + m \eta^{\alpha\beta} \delta_{m+n,0} C\, ,
\ee
where $f^{\alpha\beta\gamma}$ are the structure constants of
$o(8,24)$, and $\eta^{\alpha\beta}$ is the metric.

Let us now discuss the action of various generators of the current
algebra on the fields $M$, $\Phi$ and $G_{\mu\nu}$. We shall follow the
approach of Breitenlohner and Maison\cite{BREIT} adapted for the present
system. The central
charge $C$ simply causes a shift of $\lambda$ ({\it i.e.} a scaling of
$G_{\mu\nu}$) without any transformation of the fields $M$ and $\Phi$:
\be \label{e16b}
\delta_C G_{\mu\nu} = 2 G_{\mu\nu}, \qquad \delta_C \Phi=0, \qquad
\delta_C M=0\, ,
\ee
where for any symmetry generator $\OO$ and a field $\phi$,
$\epsilon \delta_\OO \phi$ denotes the infinitesimal transformation
of $\phi$ generated by $\OO$, $\epsilon$ being the infinitesimal
transformation parameter.
$J_0^\alpha$ generate the usual $O(8,24)$ transformations
\be \label{e16f}
\delta_0^\alpha M = -\big(T^\alpha M + M(T^\alpha)^T\big)\, , \qquad
\delta_0^\alpha \Phi =0\, , \qquad \delta_0^\alpha G_{\mu\nu}=0\, ,
\ee
where $T^\alpha$ are the generators of $o(8,24)$ algebra satisfying
\be \label{lalpha}
T^\alpha L + L (T^\alpha)^T =0\, .
\ee
The finite version of this transformation is given by,
\be \label{e16c}
M\to \Omega M \Omega^T, \qquad \Phi\to \Phi, \qquad G_{\mu\nu}\to
G_{\mu\nu},
\ee
where $\Omega$ is an O(8,24) matrix satisfying,
\be \label{omega}
\Omega L \Omega^T = L\, .
\ee

The transformations generated by $J^\alpha_n$ for $n\neq 0$ act
on $M$ in a nonlocal fashion and are more difficult to describe.
We first note that eq.\refb{e14} gives
\be \label{e17}
\rho\equiv e^{-\Phi} = \rho_+(x^+) + \rho_-(x^-)\, .
\ee
The next thing to note is that the equations of motion \refb{e15} and the
Bianchi identity of $M$,
\be \label{e21}
\p_+ (\p_- M M^{-1}) - \p_- (\p_+M M^{-1}) + [\p_- M M^{-1}, \p_+M M^{-1}]
=0\, ,
\ee
allow us to define a matrix $\U(x;v)$ satisfying\cite{NICOLAI}
\ben \label{e18}
(\p_+ \U^{-1}) \U &=& {t\over 1+t} \p_+ M M^{-1}\, , \nonumber \\
(\p_- \U^{-1}) \U &=& -{t\over 1-t} \p_- M M^{-1}\, ,
\een
\be \label{e19}
\U(x; v=0) =I_{32}\, ,
\ee
where
$v$ is an arbitrary real parameter\footnote{$v$ is related to the
spectral parameter $w$ of ref.\cite{NICOLAI} through the relation
$v=1/4w$.}, and,
\be \label{e23a}
t = {\sqrt{1+4v\rho_+} - \sqrt{1-4v\rho_-} \over \sqrt{1+4v\rho_+}
+\sqrt{1-4v\rho_-}}\, .
\ee
The Bianchi identity,
\be \label{e20}
\p_+ (\p_- \U^{-1} \U) - \p_- (\p_+\U^{-1} \U) + [\p_- \U^{-1} \U,
\p_+\U^{-1} \U] =0\, ,
\ee
follows as a consequence of eqs.\refb{e18}, \refb{e15}, \refb{e17},
\refb{e21} and \refb{e23a}.
Eq.\refb{e19} guarantees that
$\U$ has a series expansion of the form
\be \label{e22}
\U = \exp(V(x;v))\, ,
\ee
\be \label{e23}
V(x;v) = \sum_{n=1}^\infty v^{n} V^{(n)}(x)\, .
\ee

Let us now decompose $M$ as,
\be \label{e24}
M = \V \V^T\, ,
\ee
where $\V$ is an O(8,24) matrix.
The choice of $\V$ is not unique, since we can multiply $\V$ by
an arbitrary $x$ dependent $O(8)\times O(24)$ matrix from the right
without changing $M$. We shall refer to this as a {\it gauge
transformation}.  Let us now define,
\be \label{e24a}
\wh \V(x;v) = \U(x;v) \V\, .
\ee
The above procedure gives us an algorithm for constructing $\wh\V(x;v)$
for a given $M$.
On the other hand, given $\wh\V(x;v)$ which has a well defined Taylor
series expansion around $v=0$, we can define
$\V(x)$ to be $\wh \V(x;0)$, and then construct $M$
using eq.\refb{e24}. We can also reconstruct $\U$ (and hence
the potentials $V^{(n)}$) from $\wh\V$ using eq.\refb{e24a}.

Note, however, that eqs.\refb{e18} and \refb{e19} do not determine
$\U$ (and hence $\wh\V$) unambiguously for a given $M$, since we have
the freedom of multiplying
$\U$ from the left by a $v$-dependent but $x$ independent $O(8,24)$
matrix which reduces to the identity matrix at $v=0$.
This introduces an ambiguity in the definition of the potentials
$V^{(n)}$. Put another way, the equations of motion \refb{e18},
\refb{e19}, written in terms
of $\wh\U$ (or $\wh\V$), have a trivial symmetry corresponding to left
multiplication
of $\wh\U$ by a $v$-dependent $O(8,24)$ matrix, which tends to identity
as $v\to 0$. This transformation does not affect $M$.
Since $M$ was the basic variable in supergravity theory,
one might tend to conclude from
this that the extra degrees of freedom contained in the matrix $\wh\V$
due to this ambiguity are spurious. While this is certainly true for
the classical supergravity theory, this is not true in quantum string
theory. The symmetry corresponding to multiplying $\U$ from the left
by a $v$-dependent O(8,24) matrix
is broken in the quantum theory, and the full spectrum of string theory
is not invariant under this symmetry.
(One example of such a symmetry transformation would be
the translation of the would be axion in four dimensions).
Thus, in string theory, $\wh{\V}$ is a more
natural dynamical variable than $M$, and from now on we shall treat
$\wh\V$ as the basic variable of the theory. In particular, all the
transformation laws will be given for $\wh\V$ instead of $M$. (We can,
of course, find the transformation laws of $M$ from the transformation
laws of $\wh\V$, since the construction of $M$ from $\wh\V$ is free
from any ambiguity.)\footnote{On the other hand, if one is only interested
in the symmetries of the classical supergravity theory, then one can take
$M$ as the basic variable, and end up with a different symmetry
algebra\cite{SCHTWOD}.}

We are now in a position to describe the action of the generators
$J_n^\alpha$ on the basic variables in the theory, {\it e.g.}
$\Phi$, $\lambda$ and $\wh\V$.
We shall first specify the transformation laws under finite
elements of the loop group, specified by an $O(8,24)$ valued function
$g(v)$, and then specialize to infinitesimal elements.
These are given by\cite{BREIT,NICOLAI},
\be \label{efi1}
\wh\V(x;v)\to \wh\V'(x;v) = g(v) \wh\V(x;v) H(x,t)\, , \qquad
\Phi\to \Phi \, ,
\ee
where $H(x,t)$ is a suitably chosen $O(8,24)$ matrix which makes the right
hand side of eq.\refb{efi1} finite in the $v\to 0$ limit, and satisfies,
\be \label{efi2}
H^T(x, {1\over t}) H(x,t) = I_{32}\, .
\ee
Since eq.\refb{e16} determines $\lambda$ in terms of other
variables up to an overall multiplicative constant, eq.\refb{efi1}
also determines the transformation law of $\lambda$. (A more explicit
form of the transformation law of $\lambda$ has been given in
ref.\cite{BREIT} in terms of the group cocycle.)
It has been argued (see, {\it e.g.}
ref.\cite{BREIT,NICOLAI}) that these transformations preserve the
equations of motion. In other words, if we define,
\be \label{emm1}
\V'(x) = \wh \V'(x;v=0)\, ,
\ee
\be \label{emm2}
\U'(x;v) = \wh\V'(x;v) (\V'(x))^{-1}\, ,
\ee
and,
\be \label{emm3}
M'(x) = \V'(x) (\V'(x))^T\, ,
\ee
then analogs of eq.\refb{e18}, with the variables $\U$, $M$ replaced by
$\U'$, $M'$, are satisfied.

The infinitesimal transformation laws which follow from eq.\refb{efi1}
are,
\ben \label{e24c}
\wh \V'(x;v) &\equiv&
\wh \V(x;v) + \epsilon \delta^\alpha_{-n} \wh\V(x;v)
\nonumber \\
&=& \Big(1 - \epsilon T^\alpha v^{-n}\Big) \wh
\V(x;v) H^\alpha_n(x;t) \, , \cr \cr
\delta^\alpha_{-n} \Phi &=& 0\, ,
\een
where $H^\alpha_n(x;t)$ is an appropriately chosen
infinitesimal O(8,24) matrix, satisfying,
\be \label{e24d}
(H^\alpha_n)^T(x;{1\over t}) H^\alpha_n(x;t) = I_{32}\, .
\ee
This specifies the transformation
laws of $\wh\V$ under the generators $J^\alpha_n$.
For $n\le 0$ we can take $H_n^\alpha=I_{32}$.
Thus for $n=0$, $\V'(x) = \V(x) - \epsilon T^\alpha \V(x)$.
This agrees with the transformation laws given in eq.\refb{e16f}.
For $n<0$, $\wh\V'(x;v=0)=\wh \V(x;v=0)$, which shows that
the matrix $M$ does not transform under these transformations.
However, using the expansion \refb{e22}, \refb{e23} of $\U$ we see
that the potentials $V^{(m)}$ for $m\ge -n$ do transform under the
generators $J^\alpha_{-n}$.

Let us now start from the same ten dimensional theory, and
carry out the dimensional reduction in two stages.
We first compactify the directions 3-9 so that we get a three dimensional
theory. At the next stage we compactify the second direction to get a
two dimensional theory. For the first stage of dimensional reduction we
introduce coordinates
\ben \label{defbarc}
z^\bmu &=& \bar x^\bmu, \qquad 0\le \bmu \le 2, \nonumber \\
z^{\bm+2} &=& \bar y^\bm, \qquad 1\le \bm \le 7.
\een
We also define new fields:
\ben \label{e27}
&& \wh \bG_{\bm\bn}  = G^\ten_{\bm+2,\bn+2}, \quad  \wh \bB_{\bm\bn}  =
B^\ten_{\bm+2, \bn+2}, \quad  \wh \bA^I_\bm  = A^{\ten I}_{\bm+2},
\nonumber \\
&& \bA^{(2\bm-1)}_\bmu  = {1\over 2}\wh \bG^{\bm\bn} G^\ten_{\bn+2,\bmu},
\quad \bA^{(I+14)}_\bmu = -({1\over 2} A^{\ten I}_\bmu - \wh \bA^I_\bn
\bA^{(2\bn-1)}_\bmu), \nonumber \\
&&  \bA^{(2\bm)}_\bmu = {1\over 2}
B^\ten_{(\bm+2)\bmu} - \wh \bB_{\bm\bn} \bA^{(2\bn-1)}_\bmu
+ {1\over 2}\wh \bA^I_\bm \bA^{(I+14)}_\bmu, \nonumber \\
&& \bG_{\bmu\bnu} = G^\ten_{\bmu\bnu} - G^\ten_{(\bm+2)\bmu}
G^\ten_{(\bn+2)\bnu} \wh \bG^{\bm\bn}, \nonumber \\
&& \bB_{\bmu\bnu} = B^\ten_{\bmu\bnu} - 4\wh \bB_{\bm\bn}
\bA^{(2\bm-1)}_\bmu \bA^{(2\bn-1)}_\bnu - 2 (\bA^{(2\bm-1)}_\bmu
\bA^{(2\bm)}_\bnu - \bA^{(2\bm-1)}_\bnu \bA^{(2\bm)}_\bmu) \nonumber \\
&& \qquad \qquad
+ 2 \wh \bA^I_\bn (\bA^{(2\bn-1)}_\bmu \bA^{(I+14)}_\bnu
- \bA^{(2\bn-1)}_\bnu \bA^{(I+14)}_\bmu ) \, ,
\nonumber \\
&& \bPhi = \Phi^\ten - {1\over 2} \ln\det \wh \bG, \quad \quad
\quad 1\le \bm, \bn \le 7, \quad
0\le \bmu, \bnu \le 2, \quad 1\le I \le 16. \nonumber \\
\een
Here $\wh \bG^{\bm\bn}$ denotes the inverse of the matrix
$\wh \bG_{\bm\bn}$.  We also define,
\be \label{e28}
\wh \bC_{\bm\bn} = {1\over 2} \wh \bA^I_\bm \wh \bA^I_\bn\, .
\ee
For every $\bm,\bn$ ($1\le \bm,\bn \le 7$) we define $\bH_{\bm\bn}$
to be the $2\times 2$ matrix
\be \label{e29}
\bH_{\bm\bn} = \pmatrix{\wh \bG^{\bm\bn} & \wh \bG^{\bm\bp}(\wh
\bB_{\bp\bn} + \wh \bC_{\bp\bn})\cr\cr
(-\wh \bB_{\bm\bp} + \wh \bC_{\bm\bp}) \wh \bG^{\bp\bn} & (\wh \bG -
\wh \bB + \wh \bC)_{\bm\bp}
\wh \bG^{\bp\bq} (\wh \bG + \wh \bB + \wh \bC)_{\bq\bn} \cr}\, ,
\ee
and for every $\bm$, $I$ ($1\le \bm\le 7$, $1\le I\le 16$) we define
$\bQ^{(I)}_\bm$ to be the two dimensional column vector
\be \label{e30}
\bQ^{(I)}_\bm = \pmatrix{\wh \bG^{\bm\bn} \wh \bA^I_\bn \cr\cr (
\wh \bG - \wh \bB + \wh \bC)_{\bm\bn}
\wh \bG^{\bn\bp} \wh \bA^I_\bp \cr}\, .
\ee
We also define $\bK$ to be a $16\times 16$ matrix whose components are,
\be \label{e31}
\bK_{IJ} = \delta_{IJ} + \wh \bA^I_\bm \wh \bG^{\bm\bn} \wh
\bA^J_\bn \, .  \ee
In terms of $\bH$, $\bQ$ and $\bK$, we now define a $30\times 30$
matrix $\bM$ as follows:
\be \label{e32}
\bM = \pmatrix{
\bH_{11} & \cdot & \bH_{17} & \bQ^{(1)}_1 & \cdot & \bQ^{(16)}_1 \cr
  \cdot  & \cdot & \cdot  & \cdot     & \cdot & \cdot      \cr
\bH_{71} & \cdot & \bH_{77} & \bQ^{(1)}_7 & \cdot & \bQ^{(16)}_7 \cr\cr
\bQ^{(1)T}_1 & \cdot & \bQ^{(1) T}_7 &      &       &            \cr
     \cdot   & \cdot & \cdot  &           &   \bK   &            \cr
\bQ^{(16) T}_1 & \cdot & \bQ^{(16) T}_7 &     &       &            \cr
}\, .
\ee
$\bM$ satisfies
\be \label{e33}
\bM^T = \bM, \qquad \bM \bL \bM^T = \bL,
\ee
where $\bL$ is the $30\times 30$ matrix:
\be \label{eneweqn}
\bL = \pmatrix{\sigma_1 &&&& \cr &\cdot &&& \cr && \cdot && \cr
&&& \sigma_1 & \cr
&&&& -I_{16}\cr}\, .
\ee

Let us now define
\be \label{e34}
\bg_{\bmu\bnu} = e^{-2\bPhi} \bG_{\bmu\bnu}\, ,
\ee
and note that the equations of motion for the gauge fields
$\bA^{(\ba)}_\bmu$ ($1\le \ba \le 30$) allows us to define a set of
scalar fields $\bpsi^\ba$ through the relations\cite{THREED}:
\be \label{e35}
\sqrt{-\bg} e^{-2\bPhi} \bg^{\bmu \bmu'} \bg^{\bnu\bnu'} (\bM\bL)_{\ba
\bb} \bF^{(\bb)}_{\bmu'\bnu'} = {1\over 2} \epsilon^{\bmu\bnu\brho}
\p_\brho\bpsi^\ba\, .
\ee
We now define
\be \label{e37}
\wt M = \pmatrix{e^{-2\bPhi} + \bpsi^T\bL\bM\bL \bpsi & -{1\over 2}
e^{2\bPhi}\bpsi^T \bL \bpsi & \bpsi^T \bL \bM +{1\over 2} e^{2\bPhi}
\bpsi^T (\bpsi^T \bL\bpsi) \cr
+{1\over 4} e^{2\bPhi}(\bpsi^T \bL \bpsi)^2 && \cr \cr
-{1\over 2} e^{2\bPhi} \bpsi^T \bL \bpsi & e^{2\bPhi} & - e^{2\bPhi}
\bpsi^T \cr\cr
\bM\bL\bpsi +{1\over 2} e^{2\bPhi} \bpsi (\bpsi^T \bL \bpsi)
& - e^{2\bPhi} \bpsi & \bM + e^{2\bPhi} \bpsi \bpsi^T \cr}\, .
\ee
$\wt M$ satisfies
\be \label{37a}
\wt M^T = \wt M, \qquad \wt M^T L \wt M = L\, .
\ee
All information about the scalar fields $\{\bG, \bB, \bA, \bPhi\}$, as
well as the gauge fields $\{\bA^{(\ba)}_\mu\}$ (or, equivalently,
the scalar fields $\bpsi^\ba$) are contained in the matrix $\wt M$.
Finally, since in three dimensions the anti-symmetric tensor field
$\bB_{\bmu\bnu}$ has no dynamics, we can set the corresponding field
strength $\bH_{\bmu\bnu\brho}$ to zero.

The resulting equations of
motion are derivable from an action\cite{THREED}:
\be \label{e38}
S={1\over 4} \int d^3 \bar x \sqrt{-\bg} [R_{\bg} +{1\over 8}
\bg^{\bmu\bnu} Tr(\p_\bmu \wt M L \p_\bnu \wt M L)] \, .
\ee

We now compactify the direction $\bar x^2$ on a circle of length
$8\pi$ and ignore dependence of all the fields on the coordinate
$\bar x^2$.
We then get a two dimensional theory. The components $\bg_{2\mu}$ act
as gauge fields in this two dimensional theory, but since in two
dimension gauge fields have no dynamics, we can set these fields to zero.
Thus the metric $\bg_{\bmu\bnu}$ has the form:
\be \label{e39}
\bg =\pmatrix{g_{\mu\nu} & \cr & \bg_{22}\cr}\, .
\ee
The resulting action can be written as\cite{GIBBBR},
\be \label{e40}
S = 2\pi \int d^2 x \sqrt{-g} \sqrt{\bg_{22}} \, [R_g +{1\over 8}
g^{\mu\nu} Tr(\p_\mu \wt M L \p_\nu \wt M L)]\, .
\ee
Comparing the definition of $\Phi$ given in eq.\refb{e3} and the
definition of $\bg_{22}$ given in eqs.\refb{e27} and \refb{e34},
we see that,
\be \label{e40a}
\bg_{22} = e^{-2\Phi}\, .
\ee
We now write down the independent equations of motion in the gauge
$g_{\mu\nu}=\wt \lambda \eta_{\mu\nu}$:
\be \label{e40b}
\eta^{\mu\nu} \p_\mu \p_\nu (e^{-\Phi}) =0\, ,
\ee
\be \label{e40c}
\p_\mu (e^{-\Phi} \eta^{\mu\nu} \p_\nu \wt M \wt M^{-1}) =0 \, ,
\ee
and,
\be \label{e40d}
\p_\pm (\ln \wt \lambda) \p_\pm (e^{-\Phi}) =
\p_\pm^2(e^{-\Phi}) -{1\over 4} e^{-\Phi}
Tr(\p_\pm \wt M L \p_\pm \wt M L) \, .
\ee
These equations are identical to eqs.\refb{e14}-\refb{e16} with the
replacement $M\to \wt M$, $\lambda\to \wt\lambda$. Thus we can generate
another current algebra $\wh{o(8,24)}_{(2)}$ acting on these fields,
generated by the generators $\wt J^\alpha_m$:
\be \label{e40e}
[\wt J^\alpha_m, \wt J^\beta_n] = f^{\alpha\beta\gamma}
\eta_{\gamma\delta} \wt J^\delta_{m+n} + m \eta^{\alpha\beta}
\delta_{m+n,0} \wt C\, .  \ee
The subscript $(2)$ denotes the fact that in identifying the algebra
$\wh{o(8,24)}_\two$ we have given special treatment to the coordinate
$z^2$.
The action of $\wt C$, $\wt J^\alpha_m$ on various fields are defined
in a manner exactly analogous to the action of $C$, $J^\alpha_m$ on the
various fields. First of all, $\wt C$ generates a scaling of $g_{\mu\nu}$
leaving the fields $\wt M$ and $\Phi$ fixed:
\be \label{e48}
\delta_{\wt C} g_{\mu\nu}=2g_{\mu\nu}, \qquad \delta_{\wt C} \Phi=0,
\qquad \delta_{\wt C} \wt M=0\, .
\ee
In order to define an action of $\wt J^\alpha_n$ on various fields, we
define $\wt\U(x;v)$ through the equations:
\ben \label{e49}
(\p_+ \wt\U^{-1}) \wt\U &=& {t\over 1+t} \p_+ \wt M \wt M^{-1}\, ,
\nonumber \\
(\p_- \wt \U^{-1}) \wt \U &=& - {t\over 1-t} \p_- \wt M \wt M^{-1}\, ,
\een
\be \label{e50}
\wt \U(x; v=0) =I_{32}\, ,
\ee
where $t$ has been defined in eq.\refb{e23a}.
$\wt\U$ has a series expansion in $v$ of the form
\be \label{e51}
\wt\U = \exp(\wt V(x;v))\, ,
\ee
\be \label{e52}
\wt V(x;v) = \sum_{n=1}^\infty v^{n}
\wt V^{(n)}(x)\, .
\ee

We now decompose $\wt M$ as,
\be \label{e53}
\wt M = \wt \V \wt \V^T\, ,
\ee
where $\wt\V$ is an O(8,24) matrix. We also define
\be \label{e54}
\wh {\wt \V}(x;v) = \wt \U(x;v) \wt \V\, .
\ee
As for the $J^\alpha_n$'s, the transformation laws of the field
$\wh{\wt \V}$ under $\wt J^\alpha_n$ are defined as follows:
\ben \label{e56}
\wh {\wt \V}''(x;v) &\equiv&
\wh{\wt \V}(x;v)
+\epsilon \wt\delta^\alpha_{-n} \wh{\wt \V}(x;v) \nonumber \\
&=& \big(1 - \epsilon T^\alpha v^{-n}\big) \wh {\wt \V}(x;v) \wt
H^\alpha_n(x;t) \, ,
\een
where $\wt H^\alpha_n(x;t)$ is a suitably chosen infinitesimal
$O(8,24)$ matrix  which makes the right hand side of eq.\refb{e56}
finite in the $v\to 0$ limit, and satisfies,
\be \label{e57}
(\wt H^\alpha_n)^T(x;{1\over t}) \wt H^\alpha_n(x;t) = I_{32}\, .
\ee
This specifies the transformation
laws of $\wh{\wt\V}$ under the generators $\wt J^\alpha_n$.
The finite form of this transformation is,
\be \label{efi3}
\wh{\wt\V}''(x;v) = \wt g(v) \wh{\wt\V}(x;v) \wt H(x,t)\, .
\ee
$\Phi$ remains invariant under these transformations, and the
transformation law of $\wt\lambda$ is determined using the equation
of motion \refb{e40d}.

Since the symmetry algebras $\wh{o(8,24)}$ and
$\wh{o(8,24)}_{(2)}$ are in fact the symmetry algebras of the same theory,
the two algebras must be related by an automorphism. We now
ask:

\noindent {\bf Question: What is the automorphism that relates the algebras
$\wh{o(8,24)}$ and $\wh{o(8,24)}_\two$?}

\noindent {\bf Answer:} Let $T^0$ be the $o(1,1)$ generator
\be \label{e45}
T^0 = \pmatrix{1&& \cr & -1 & \cr && 0_{30}\cr }\, .
\ee
We arrange the generators $T^\alpha$ of $o(8,24)$ such that,
\be \label{e46}
[T^0, T^\alpha] = \lambda^\alpha T^\alpha\, ,
\ee
where $\lambda^\alpha$ are numbers.
Then the generators $C$, $J_n^\alpha$ of $\wh{o(8,24)}$
and the generators $\wt C$,
$\wt J_n^\alpha$ of $\wh{o(8,24)}_\two$ are related by
the following automorphism:
\be \label{e47b}
\wt C = C,
\ee
\be \label{e47a}
\wt J_n^\alpha = J_{n+\lambda^\alpha}^\alpha +
\delta_{\alpha,0} \, \delta_{n,0} \, C\, .
\ee

\noindent {\bf Proof}:
First of all, both $C$ and $\wt C$ generate a scaling of the components
$G^\ten_{\mu\nu}$ ($0\le \mu, \nu \le 1$) of the ten dimensional
metric leaving all other components of the ten dimensional fields
fixed. Thus eq.\refb{e47b} is manifest.

For proving eq.\refb{e47a} we follow the approach of ref.\cite{BREIT}.
In appendix A we have proved the following relation between $\wh\V$ and
$\wh{\wt\V}$:
\be \label{e61}
\wh{\wt \V}(x;v) = (v)^{- T^0} \wh \V(x;v) h(x;t)\, ,
\ee
where $h(x,t)$ is an appropriately chosen
O(8,24) transformation satisfying
\be \label{e62}
h^T(x,{1\over t}) h(x,t) = I_{32}\, .
\ee
Eq.\refb{e61} implies that if a given element of the loop group
$\wh{O(8,24)}$ acts on $\wh{\wt\V}$ as,
\be \label{exx1}
\wh{\wt\V}(x;v)\to \wt g(v) \wh{\wt\V}(x;v) \wt H(x,t)\, ,
\ee
then it acts on $\wh\V$ as,
\be \label{exx2}
\wh{\V}(x;v)\to g(v) \wh{\V}(x;v) H(x,t)\, ,
\ee
where,
\be \label{exx3}
g(v) = (v)^{T^0} \wt g(v) (v)^{-T^0}\, .
\ee
The infinitesimal version of the above equation will give us the
relations between $J^\alpha_n$ and $\wt J^\alpha_n$.
{}From eqs.\refb{e24c}, \refb{e61}, we get,
\ben \label{e63}
\wh{\wt \V}'(x;v) &\equiv& (v)^{-T^0} \wh \V'(x;v) h'(x,t) \nonumber \\
& = & (v)^{-T^0} (1 - \epsilon T^\alpha v^{-n}) \wh{\V}(x;v)
H^\alpha_n(x,t) h'(x,t) \nonumber \\
& = & (v)^{-T^0} (1 - \epsilon
T^\alpha v^{-n}) (v)^{T^0} \wh{\wt \V}(x;v)
h^{-1}(x;t) H^\alpha_n(x,t) h'(x,t) \nonumber \\
& = & (1 - \epsilon T^\alpha (v)^{-n-\lambda^\alpha}) \wh{\wt \V}(x;v)
h^{-1}(x,t) H_n^\alpha(x,t) h'(x,t)\, . \nonumber \\
\een
Comparing this with eq.\refb{e56} we see that this is the transformation
generated by $\wt J^\alpha_{-n-\lambda^\alpha}$. Thus we get
\be \label{e64}
\delta^\alpha_{-n}\wh\V = \wt\delta^\alpha_{-n-\lambda^\alpha} \wh\V\, .
\ee
This proves eq.\refb{e47a} acting on the field $\wh\V$
(and hence $\wh{\wt\V}$). Note that
the central charge term appearing on the right hand side of eq.\refb{e47a}
does not show up here, since it does not act on $\wh\V$.

To see the presence of the central charge term in eq.\refb{e47a}, we
need to compare the action of $J^0_0$ and $\wt J^0_0$ on the component
$G^\ten_{\mu\nu}$ of the ten dimensional metric.
This can be easily done, since we know their action on the fields
$M$, $\Phi$ and $G_{\mu\nu}$ ($\wt M$, $\Phi$, $g_{\mu\nu}$).
The non-trivial transformations are,
\ben \label{ecomp1}
\delta_0^0 G^\ten_{22} & = & 2 G^\ten_{22}, \qquad
\delta^0_0 G^\ten_{2,\bm+2}
= G^\ten_{2,\bm+2}, \qquad \delta^0_0 B^\ten_{2, \bm+2}
= B^\ten_{2, \bm+2}, \cr \cr
\delta^0_0 A^{\ten I}_2 & = & A^{\ten I}_2, \qquad \delta^0_0 \Phi^\ten =1,
\qquad \delta^0_0 G^\ten_{\mu\nu}=0, \cr\cr
&& \qquad 1\le \bm \le 7, \qquad 1\le I\le 16, \qquad 0\le \mu,\nu
\le 1\, ,
\een
\ben \label{ecomp2}
\wt \delta_0^0 G^\ten_{22} & = & 2 G^\ten_{22}, \qquad \wt \delta^0_0
G^\ten_{2,\bm+2}
= G^\ten_{2,\bm+2}, \qquad \wt \delta^0_0 B^\ten_{2, \bm+2}
= B^\ten_{2, \bm+2}, \cr \cr
\wt \delta^0_0 A^{\ten I}_2 & = & A^{\ten I}_2, \qquad
\wt \delta^0_0 \Phi^\ten =1,
\qquad \wt \delta^0_0 G^\ten_{\mu\nu}=2 G^\ten_{\mu\nu}. \nonumber \\
\een
Thus the difference between these two transformations is
simply a scaling of $G^\ten_{\mu\nu}$. The result can be written in
terms of the four dimensional fields as,
\ben \label{e77}
&& \wt \delta^0_0  M - \delta^0_0 M=0, \qquad \wt \delta^0_0 \Phi
-\delta^0_0\Phi=0, \qquad
\cr\cr
&& \wt \delta^0_0 G_{\mu\nu} - \delta^0_0 G_{\mu\nu} = 2 G_{\mu\nu} ,
\qquad 0\le \mu,\nu \le 1\, .
\een
Comparing this with eq.\refb{e16b} we see that
\be \label{e78}
\wt J^0_0 - J^0_0 = C\, .
\ee
This agrees with eq.\refb{e47a} for $n=0$, $\alpha=0$.

\sectiono{Discrete Duality Symmetry Group of the Two Dimensional
String Theory}

In the previous section we have seen that the equations of motion
of the massless fields in the two dimensional string theory have
an $\wh{O(8,24)}$ loop group symmetry.
The question that we shall be addressing in this section is: what
(discrete) subgroup
of this infinite dimensional group is a symmetry of the
full string theory?
First of all, a discrete subgroup $O(8,24;Z)$ is already known
to be a symmetry of the full string theory. This is the $T$-duality
group of the two dimensional theory. Besides this, if we regard the
two dimensional string theory to be the result of compactification of
a four dimensional theory, then the four dimensional theory is
expected to be invariant under an $S$-duality symmetry group $SL(2,Z)$.
This can also be identified as a subgroup of the loop group
$\wh{O(8,24)}$.
If we believe that compactification does not destroy $S$-duality
invariance of a theory, then this $SL(2,Z)$ subgroup of the loop
group must also be a symmetry of the full two dimensional string
theory. Since the $S$-duality group and the $T$-duality group in
two dimensions do not commute, they will generate a much bigger
subgroup of $\wh{O(8,24)}$. This, we believe, will be the minimal
duality symmetry group of the two dimensional string theory.

At this point, we should note that there are many (precisely 28)
different ways in
which a two dimensional string theory may be regarded as a compactified
four dimensional theory, giving rise to many different $S$-duality
groups. These different choices will correspond to
the choice of two out of eight coordinates $z^2,\ldots z^9$
that we use to label the
non-compact directions of the four dimensional theory. However,
these different choices are permuted by the $T$-duality group,
and hence the final duality group generated by the
$T$-duality group $O(8,24;Z)$ and the $S$-duality group $SL(2,Z)$
will be
independent of the initial choice of the $SL(2,Z)$ group.

Thus, in order to describe the duality group of the two dimensional
theory, we should first determine how the $S$- and the
$T$-duality groups are embedded in the loop group $\wh{O(8,24)}$.
This is done by specifying the $O(8,24)$ group valued function $g(v)$
for each generator of the $S$- and the $T$- duality groups.
This representation, however,
is not faithful, in that
a given $g(v)$ will represent a whole one dimensional orbit in the
$\wh{O(8,24)}$ group which differ from each other by the action of
the central charge. We shall later discuss how this
ambiguity gets resolved.

We shall first study how the O(8,24;Z) target space duality
transformations are embedded in $\wh{O(8,24)}$.  This is easy,
since this O(8,24;Z) is a subgroup of the $O(8,24)$ group that is
generated
by the generators $J_0^\alpha$. If $U$ is an $O(8,24;Z)$
matrix, its action on $\wh\V$ is given by,
\be \label{eyy1}
\wh\V\to U\wh\V\, .
\ee
Comparing with eq.\refb{efi1} we see that this corresponds to a loop
group element,
\be\label{en1}
g(v; U\in O(8,24;Z)) = U\, .
\ee
At this stage, let us briefly
recall the way we characterize the $O(8,24;Z)$ matrices $U$.
A general $O(8,24)$ group element
acts naturally on a 32 dimensional vector space.
We define a lattice $\Lambda_{32}$ in this vector space, spanned by
vectors of the form:
\be \label{seight}
\pmatrix{\vec m \cr \vec \alpha\cr}\, ,
\ee
where $\vec m$ is a 16 dimensional vector with integer entries,
and $\vec\alpha$
is a 16 dimensional vector in the $E_8\times E_8$ root lattice. The
O(8,24;Z) subgroup of O(8,24) is defined to be the group of O(8,24)
matrices that preserve the lattice $\Lambda_{32}$.

Let us now see how the $S$-duality transformations fit inside the
loop group $\wh{O(8,24)}$. We shall consider the $S$-duality group
of the four dimensional theory obtained by compactifying the
directions $z^4-z^9$. This $S$-duality group has a natural embedding
in the group $O(8,24)_{(2)}$ generated by $\wt J^\alpha_0$.
If $\wt U$ is an $O(8,24;Z)_{(2)}$ matrix, then its action on
$\wh{\wt\V}$ is given by,
\be \label{eyy2}
\wh{\wt\V}\to \wt U \wh{\wt\V}\, .
\ee
Comparing with \refb{efi3} we see that this corresponds to the loop
group element
\be \label{eyy3}
\wt g(v;\wt U\in O(8,24;Z)_{(2)}) = \wt U\, .
\ee
Eq.\refb{exx3} now gives,
\be \label{eyy4}
g(v;\wt U\in O(8,24;Z)_{(2)}) = (v)^{T^0} \wt  U (v)^{-T^0}\, .
\ee
Using eq.\refb{eyy4}, we can construct the loop group element $g(v)$
corresponding to an $S$-duality transformation.
A generic $SL(2;Z)$ transformation is represented by an $O(8,24;Z)_{(2)}$
matrix of the form:\footnote{This is obtained by appropriate
rearrangements of the rows and columns in the expression given in
ref.\cite{THREED}.}
\be \label{ea1}
\wt U(p,q,r,s)=
\pmatrix{ p & 0 & 0 & -q & \cr 0 & s & r & 0 & \cr 0 & q & p & 0 & \cr
-r & 0 & 0 & s & \cr &&&& I_{28}\cr},
\ee
where $p$, $q$, $r$, $s$ are integers satisfying
\be \label{ea1n}
ps - qr = 1\, .
\ee
This discrete group is generated by the following two $SL(2,Z)$
transformations:
\ben \label{ea2}
\T_{(23)}  & : & \wt U =
\pmatrix{ 1 & 0 & 0 & -1 & \cr 0 & 1 & 0 & 0 & \cr 0 & 1 & 1 & 0 & \cr
0 & 0 & 0 & 1 & \cr &&&& I_{28}\cr}, \nonumber \\
\nonumber \\
\SS_{(23)} & : & \wt U =
\pmatrix{ 0 & 0 & 0 & -1 & \cr 0 & 0 & -1 & 0 & \cr 0 & 1 & 0 & 0 & \cr
1 & 0 & 0 & 0 & \cr &&&& I_{28}\cr}, \nonumber \\
\een
where the subscript $(23)$ is a reminder of the fact that the directions
$z^2$ and $z^3$ are given special treatment in this case.
Eq.\refb{eyy4} now shows that
\be \label{ea11}
g(v;\T_{(23)}) =
\pmatrix{ 1 & 0 & 0 & -v & \cr 0 & 1 & 0 & 0 & \cr 0 &
v & 1 & 0 & \cr
0 & 0 & 0 & 1 & \cr &&&& I_{28}\cr},
\ee
and,
\be \label{ea12}
g(v;\SS_{(23)}) =
\pmatrix{ 0 & 0 & 0 & -v & \cr 0 & 0 & -v^{-1} & 0 & \cr 0 &
v & 0 & 0 & \cr
v^{-1} & 0 & 0 & 0 & \cr &&&& I_{28}\cr}.
\ee
Let us define by $G$ the full discrete subgroup of $\wh{O(8,24)}$,
generated by $g(v; \T_{(23)})$, $g(v; \SS_{(23)})$,
and the $T$-duality group elements $g(v; U\in O(8,24;Z))$
defined in eqs.\refb{ea11}, \refb{ea12} and \refb{en1} respectively.
This is the minimal duality symmetry group of the two
dimensional string theory.
As already mentioned, the generators of the other $S$-duality
groups can be obtained as combinations  of
$g(v; \T_{(23)})$, $g(v; \SS_{(23)})$, and $g(v; U\in O(8,24;Z))$.
In general, the $S$-duality generators
$g(v; \T_{(ij)})$ and $g(v; \SS_{(ij)})$
are obtained from $g(v; \T_{(23)})$ and $g(v; \SS_{(23)})$
by the following rearrangement of the rows and columns:
\be \label{ea13}
(1,2) \to (2i-3, 2i-2), \qquad (3,4) \to (2j-3, 2j-2)\, .
\ee

In order to get some further insight into the structure of the
group $G$, let us define $\wh{O(8,24;Z)}$ to be the following subgroup
of $\wh{O(8,24)}$. $g(v)\in \wh{O(8,24;Z)}$ if,
\be \label{edis1}
g(v) L g(v)^T = L\, ,
\ee
and $g(v)$ has an expansion of the form:
\be \label{edis2}
g(v) = \sum_{n=-\infty}^\infty v^{-n} g_n\, ,
\ee
where each $g_n$, acting on a vector in the
lattice $\Lambda_{32}$, gives another vector in $\Lambda_{32}$.
{}From eqs.\refb{ea11}, \refb{ea12} and \refb{en1} we can easily
verify that each of the generators of the duality group $G$ is an
element of $\wh{O(8,24;Z)}$. Hence $G$ itself must be a subgroup of
$\wh{O(8,24;Z)}$.
$G$, however, does not contain all the elements of $\wh{O(8,24;Z)}$.
In particular, it does not contain the element represented by the
$O(8,24)$ valued function $(v)^{T^0}$. This can be seen from the fact
that for all the generators of $G$, $g(v=1) g^{-1}(v=-1)$
represents an $O(8,24)$ matrix which can be continuously deformed to
the identity element. This is not the case for the matrix
$(-1)^{T^0}$.  It is still tempting to conjecture\cite{HULL}
that the full duality symmetry group of the theory is
$\wh{O(8,24;Z)}$, but justifying this will require further work.
In particular, we need to study if $\wh{O(8,24;Z)}$
elements like $(v)^{T^0}$ generate duality symmetries of the two
dimensional string theory.

Finally we turn to the complications that arise due to the presence of
the central charge. As we have already pointed out, a given
$O(8,24)$ valued function $g(v)$
actually represents
to a one parameter family of $\wh{O(8,24)}$ group elements, related by
the action of the central charge. Thus one needs to understand how for
a given $g(v)$, representing an element of
the group $G$, one reconstructs a specific
element of the $\wh{O(8,24)}$ group. This can be done by representing
$g(v)$ as a product of
$g(v; \T_{(23)})$, $g(v; \SS_{(23)})$,
and the $T$-duality group elements $g(v; U\in O(8,24;Z))$.
For each of these elements,
we know the transformation laws of various fields, including the two
dimensional metric. Thus once we express an element of $G$ as a product
of these transformations, we know how that element of $G$ acts on
various fields. This determines the specific element of $\wh{O(8,24)}$
that a given $G$ transformation corresponds to.

There of course remains the question as to whether two different ways
of expressing a given $g(v)$ as products of
$g(v; \T_{(23)})$, $g(v; \SS_{(23)})$,
and $g(v; U\in O(8,24;Z))$ may lead to two different
transformation laws of the two dimensional metric. If this is the case,
then, by following one series of transformations by the inverse of the
other, we can get an element of the duality group which would correspond
to scaling the two dimensional metric by some fixed amount without
affecting any other field. If this had been a symmetry of the full
string theory, it would have been valid order by order in perturbation
theory, since this transformation does not act on the string coupling
constant. In other words, this should have been part of the $T$-duality
transformation of the theory. We know, however, that the $T$-duality
group does not contain any such element. This, in turn, implies that
our initial assumption must be wrong, and two different ways of
expressing a specific matrix representing an element of $G$ must
lead to identical transformations of the two dimensional metric.
Thus, although the representation that we have chosen is not a
faithful representation of the continuous group $\wh{O(8,24)}$, it
provides us with a faithful representation of the discrete duality
symmetry group $G$ of the two dimensional string theory.

\sectiono{Conclusion}

In this paper we have analysed the structure of the discrete duality
symmetry group of the two dimensional string theory, obtained by
compactifying the heterotic string theory on an eight dimensional
torus. The minimal duality group $G$, generated by the known $S$-
and $T$-duality transformations of the theory,
turns out to be a discrete subgroup of the loop
group $\wh{O(8,24)}$. The generators of this discrete group are
represented by the $O(8,24)$ valued functions given in
eqs.\refb{ea11}, \refb{ea12} and \refb{en1}. This discrete group is
a subgroup of $\wh{O(8,24;Z)}$, $-$ the group of $O(8,24)$ valued
functions of a variable $v$ satisfying the criteria that if we
expand the matrix valued function in positive and negative powers of $v$,
then each term in this power series expansion, acting on an element
of the even,
self-dual, Lorentzian lattice $\Lambda_{32}$ defined in eq.\refb{seight},
will give another element of $\Lambda_{32}$.
At present it remains an open question whether the theory is invariant
under the full $\wh{O(8,24;Z)}$ group.

\appendix

\sectiono{Relation between $\wh \V$ and $\wh {\wt \V}$}

In this appendix we prove eq.\refb{e61}. For this, let us define,
\be \label{e101}
\wh \V_1 \equiv (v)^{-T^0} \wh \V(x;v) h(x,t)\, .
\ee
Thus we need to prove that $\wh {\wt \V}=\wh \V_1$. First we note from
eq.\refb{efi1} and \refb{e101} that $\wh \V_1$ and $\wh \V$ are related
by a finite $\wh{O(8,24)}$ loop group transformation. Since this
transformation is a symmetry of the equations of motion, $\wh \V_1$
and $\wh \V$ satisfy the same equations of motion. In other words, we
have,
\be \label{e102}
\wh \V_1 = \wh \U_1(x;v) \V_1(x)\, ,
\ee
\be \label{e103}
\V_1(x) \equiv \wh \V_1(x;v=0)\, ,
\ee
with $\wh \U_1$ satisfying the dynamical equations:
\ben \label{e104}
(\p_+ \U_1^{-1}) \U_1 &=& {t\over 1+t} \p_+M_1 (M_1)^{-1}\, , \nonumber \\
(\p_- \U_1^{-1}) \U_1 &=& -{t\over 1-t} \p_-M_1 (M_1)^{-1}\, ,
\een
where,
\be \label{e105a}
M_1 = \V_1 \V_1^T\, .
\ee
We furthermore note that up to a gauge transformation, and the ambiguity
mentioned in the text, $\wh \V_1$ is
determined in terms of $M_1$ in the same way that $\wh \V$ ($\wh{\wt\V}$)
is determined
in terms of $M$ ($\wt M$).
Comparing the above equations with
eqs.\refb{e49}, \refb{e50}, \refb{e53} and \refb{e54} we see that
$\wh\V_1$ satisfies the defining equations of $\wh{\wt\V}$, provided,
\be \label{e106}
M_1=\wt M\, .
\ee
Thus the proof of eq.\refb{e61} boils down to proving
eq.\refb{e106}.\footnote{Note that the construction of $\wh{\wt\V}$ from
$\wt M$ using eqs.\refb{e49}, \refb{e50}, \refb{e53} and \refb{e54}
suffers from ambiguities similar to those appearing in the construction of
$\wh\V$ from $M$. Thus even if we prove eq.\refb{e106}, it will only
imply that $\wh{\wt\V}$ is equal to
$\wh{\V}_1$ given in eq.\refb{e101} up to this ambiguity
(and gauge transformations). But we shall take the stand that once
eq.\refb{e106} is proved, we can take eq.\refb{e61} to be the
definition of $\wh{\wt\V}$. What this means is that given
$\wh\V$, eq.\refb{e61} gives a procedure for constructing $\wh{\wt\V}$
satisfying eqs.\refb{e49}, \refb{e50}, \refb{e53} and \refb{e54}.}

We shall carry out the proof of eq.\refb{e106} in three stages.
In the first stage, we shall construct $\wh\V$ to required order in $v$
using eqs.\refb{e18}, \refb{e19}, \refb{e24}  and \refb{e24a}.
{}From this we construct $\wh\V_1$ using eq.\refb{e101}.
Finally we shall construct $M_1$ using eq.\refb{e105a} and
compare with the expression for $\wt M$ given in eq.\refb{e37}.

We begin with the construction of $\wh\V$.
This involves two steps, construction of an O(8,24) matrix $\V$ satisfying
eq.\refb{e24}, and the construction of $\U$ from eqs.\refb{e18},
\refb{e19}.
For this purpose it will be convenient to use an expression for the
matrix $M$ in terms of the fields introduced in
eqs.\refb{e27}-\refb{e32}. It is given by,
\be \label{em1}
M = \pmatrix{
(\bG_{22})^{-1} & -{1\over 2} (\bG_{22})^{-1} \bchi^T \bL \bchi &
- (\bG_{22})^{-1} \bchi^T \cr\cr
-{1\over 2} (\bG_{22})^{-1}
\bchi^T \bL \bchi & \bG_{22} + \bchi^T\bL\bM\bL\bchi &
\bchi^T \bL \bM +{1\over 2} (\bG_{22})^{-1} \cr
& +{1\over 4} (\bG_{22})^{-1} (\bchi^T \bL \bchi)^2 &
\bchi^T (\bchi^T \bL\bchi)
\cr \cr
-(\bG_{22})^{-1} \bchi &
\bM\bL\bchi +{1\over 2} (\bG_{22})^{-1}\bchi (\bchi^T \bL \bchi)
& \bM + (\bG_{22})^{-1} \bchi \bchi^T \cr}\, ,
\ee
where,
\be \label{em2}
\bchi = 2 \pmatrix{\bA_2^{(1)} \cr \cdot \cr \cdot \cr \bA_2^{(30)}\cr}
\, .
\ee
It is a straightforward algebraic exercise to check that eq.\refb{em1}
agrees with the expression for $M$ given in eq.\refb{e8}.
A convenient $\V$ satisfying \refb{e24}
can now be found using the expression for
$M$ given in eq.\refb{em1}. We take,
\be \label{em8}
\V =\pmatrix{\V_{11} & 0 & 0 & \cdot & 0 \cr
\V_{21} & \V_{22} & \V_{23} & \cdot & \V_{2,32} \cr
\V_{31} & 0       &         &       &           \cr
\cdot   & \cdot   &         & \bar\V &          \cr
\V_{32,1} & 0     &         &        &          \cr}\, ,
\ee
where
\ben \label{em3}
&& \V_{11} = (\bG_{22})^{-{1\over 2}}, \qquad \V_{22} =
(\bG_{22})^{1\over 2}, \qquad
\V_{21} = -{1\over 2} (\bG_{22})^{-{1\over 2}}
\bchi^T \bL \bchi, \nonumber \\
&& \pmatrix{\V_{31} \cr \cdot \cr \cdot \cr \V_{32,1}\cr}
= -(\bG_{22})^{-1/2}\bchi, \qquad
\pmatrix{\V_{23} \cr \cdot \cr \cdot \cr \V_{2,32}\cr}
= \bar\V^T\bL\bchi,
\een
and $\bar\V$ is an $O(7,23)$ matrix, satisfying,
\be \label{em4}
\bar \V^T \bL \bar\V = \bL, \qquad \bar\V \bar\V^T = \bM\, .
\ee

The construction of $\U$ proceeds as follows. We look for a
solution of eqs.\refb{e18} of the form \refb{e22}, \refb{e23}.
Comparing terms of order $v$ on both sides of eq.\refb{e18}
we get,
\be \label{e110}
\epsilon^{\mu\nu}\p_\nu (V^{(1)}) = \rho\eta^{\mu\nu}
(\p_\nu M M^{-1})\, ,
\ee
Using the definition \refb{e35} for the vector $\bpsi$, it is a
straightforward algebraic exercise to verify the following relations:
\be \label{e120}
\bL \bpsi = \pmatrix{ V^{(1)}_{13} \cr \cdot \cr \cdot \cr
V^{(1)}_{1,32}\cr}\, , \qquad
-\bpsi = \pmatrix{ V^{(1)}_{32} \cr \cdot \cr \cdot \cr
V^{(1)}_{32,2}\cr}\, .
\ee
Similar equations for $V^{(n)}$ for $n>1$ may be written down, but
we shall not require them. The only other information we shall
need is that since $V^{(n)}$ are the generators of the $o(8,24)$
algebra, we have
\be \label{e111a}
V^{(n)}_{12} = 0.
\ee

We can now proceed to compute $\wh\V_1$ using eq.\refb{e101}. {}From
eq.\refb{e23a} we get,
\be \label{e115}
v = t\big(\rho (1 + t^2) + 2 t \wt \rho\big)^{-1}\, ,
\ee
where $\wt\rho\equiv \rho_--\rho_+$.
Thus
\be \label{e116}
(v)^{-T^0} = \pmatrix{ t^{-1} \big( \rho ( 1 + t^2) + 2
t \wt \rho \big) && \cr & t \big(\rho (1 + t^2) +  2t \wt\rho\big)^{-1}
& \cr
&& I_{30} \cr}\, .
\ee
The construction of $\wh \V_1$ also requires the knowledge of the
$O(8,24)$ matrix $h(x,t)$ satisfying \refb{e62}. Here,
\be \label{e117}
h(x,t) = \pmatrix{t && \cr & t^{-1} & \cr && I_{30}\cr}\, .
\ee
With this choice of $h(x;t)$ the $\wh \V_1$ defined in
eq.\refb{e101} has a finite $v\to 0$ limit. In fact, using the
relation between the components of $V^{(1)}$ and $\bpsi$
mentioned above, one gets
\ben \label{e118}
\V_1(x) &\equiv& \wh \V_1(x;v=0) \nonumber \\
&& \nonumber \\
& = &
\pmatrix{
\rho \V_{11} & -{1\over 2}\rho^{-1}\V_{22}\bpsi^T\bL\bpsi  &
& \bpsi^T\bL\bar\V & \cr
0 & \rho^{-1}\V_{22} & 0 & \cdot & 0 \cr
0 & &&& \cr
\cdot & -\rho^{-1}\V_{22}\bpsi && \bar \V & \cr
0 & &&& \cr}\, . \nonumber \\
\een
Using eqs.\refb{e37}, \refb{em4} and the relations
\be \label{em9}
\rho \V_{11} = e^{-\bPhi}, \qquad \rho^{-1} \V_{22} = e^{\bPhi},
\ee
one can easily verify that
\be \label{e122}
M_1 \equiv \V_1\V_1^T = \wt M\, .
\ee
This, in turn, proves eq.\refb{e61}.

\underbar{Acknowledgement}:
I wish to thank John Schwarz for many useful discussions, for
bringing many of the relevant papers in this subject to my attention,
for his comments on the manuscript,
and for communicating his results to me prior to publication.
I would also like to thank J. Maharana, S. Mukhi, S. Shenker, C. Vafa
and S. Wadia for discussions.

\end{document}